
\def\t{T}
\def\l{L}
\def\tphi{\tilde\phi}
\def\tlambda{\tilde\lambda}
\def\wfr{\bar Z}
\def\bphi{\bar\phi}
\def\phicl{\phi_{\rm cl}}
\def\phimin{\phi_{\rm min}}
\def\co{{\cal O}}
\def\tr{{\rm Tr}}

\font\titlefont=cmbx10 scaled\magstep1
\null
\rightline{SISSA 39/95/EP}
\null
\vskip 3cm
\centerline{\titlefont THE BETA FUNCTIONS}
\centerline{\titlefont OF A SCALAR THEORY}
\centerline{\titlefont COUPLED TO GRAVITY}
\smallskip
\vskip 3cm
\centerline{\bf L. Griguolo}
\smallskip
\centerline{Center for Theoretical Physics}
\centerline{Massachusetts Institute of Technology}
\centerline{77 Massachusetts Avenue, Cambridge, MA 02139}
\centerline{and}
\centerline{Istituto Nazionale di Fisica Nucleare,
Sezione di Padova}
\bigskip
\centerline{\bf R. Percacci }
\smallskip
\centerline{International School for Advanced Studies, Trieste,
Italy}
\centerline{via Beirut 4, 34014 Trieste, Italy}
\centerline{and}
\centerline{Istituto Nazionale di Fisica Nucleare,
Sezione di Trieste}
\vskip 1.8cm
\centerline{\bf Abstract}
\smallskip
\midinsert
\narrower\narrower\noindent
We study a scalar field theory coupled to gravity
on a flat background, below Planck's energy.
Einstein's theory is treated as an effective field theory.
Within the context of Wilson's renormalization group,
we compute gravitational corrections to the beta functions
and the anomalous dimension of the scalar field, taking
into account threshold effects.
\endinsert
\bigskip
\vfil
\eject
\leftline{\bf 1. Introduction}
\smallskip
\noindent
Our present understanding of the four known interactions is based on the
Standard Model of particle physics and on General Relativity.
With the exception of astrophysical phenomena,
there seem to be few instances where the full machinery of these theories
need to be applied together. In particular, for all the situations which are
within experimental reach, the mutual influence between gravitational
and particle phenomena is very weak.
On one hand, the gravitational fields produced by the elementary
particles is exceedingly weak; on the other, the contribution of
gravitons to quantum amplitudes are suppressed by powers of momentum
over the Planck mass, and therefore are negligible at
presently available energies.

In grand unified theories, however, one has to discuss effects
occurring at energies which are quite close to the Planck scale.
In these situations it is obviously important to be able to
estimate the effects due to the gravitational field.
As a first step in this direction, we will investigate here the
influence of gravity on the renormalization group of a scalar
field.

The renormalization group describes the change in the effective
action as some external scale $k$ is varied. The physical
meaning of this parameter $k$ can vary from one problem to another,
but in all cases it acts effectively like an infrared cutoff.
This point of view originated with work of Wilson and others
[1] and has undergone considerable development recently [2].
The particular implementation of these ideas that we shall use
here is the so called average effective action, which has been
studied, in the case of scalar fields, in [3].
In this paper we will consider the average effective action of
a scalar field coupled to gravity, at energies below Planck's
energy. We will assume that in this regime
gravity can be described by Einstein's
theory, treated as an effective field theory with a cutoff
at the Planck scale.
We make two simplifying assumptions:
Newton's constant is assumed not to run, and higher derivative
terms are neglected.
Both these assumptions will probably be violated as one approaches
the Planck energy.
A physical picture of the transition from the sub-planckian
to the super-planckian regime is given in [4].

This paper is organized as follows. In sections 2 and 3 we
define the average effective action for a scalar field, paying
due attention to the overall normalization, and compute the
beta functions of some of its parameters. In sections 4 and 5 we
compute the gravitational corrections to the beta functions
of the scalar fields. In section 6 we put together our results
and draw our conclusions.
\bigskip
\leftline{\bf 2. Average effective action for a scalar field}
\smallskip
\noindent
In order to illustrate the general method in a simpler setting
and also to establish some formulae that will be useful later,
we will discuss first the case of a self-interacting scalar field
$\phi$ in the spontaneously broken phase in flat space.
The aim of this section is to give the definition of the average
effective action $\Gamma_k$, an action describing accurately the
physics of the system at momentum scale $k$. In particular, we will
be interested in the average effective potential $V_k(\phi)$
and in the ``average wave function renormalization''
$Z_k(\phi)$, which are defined by the expansion of $\Gamma_k$ in
powers of derivatives of $\phi$:
$$
\Gamma_k(\phi)=\int d^4x\,\left[V_k(\phi)
+{1\over2}Z_k(\phi)\partial_\mu\phi\partial^\mu\phi+\ldots\right]
\ .\eqno(2.1)
$$
In the next section we will write the renormalization group equation
for some of the parameters appearing in $V_k$ and $Z_k$.

Let $\Gamma(\phi)$ be the usual effective action,
which is obtained by means of a Legendre transformation,
using the background field method. At one loop it is given by
$$
\Gamma(\phi)={1\over2}
\ln\left({{\rm Det}\co(\phi)\over{\rm Det}\co(\phi_0)}\right)\ ,\eqno(2.2)
$$
where $\co(\phi)$ is the operator describing the propagation of
small fluctuations around the background $\phi$, and $\phi_0$
is a fixed constant field, that one may conveniently choose as
the minimum of the effective potential.
The denominator in (2.2) gives the correct overall normalization
of the effective action. It does not depend
on $\phi$ and therefore is often disregarded, but it will play
a role in what follows.

The determinants in (2.2) can be expanded perturbatively as sums of
one loop graphs. Each of these involves a single integration over
momenta. We assume that both integrals have been regularized by
introducing some ultraviolet cutoff $\Lambda$. This UV cutoff will not
appear explicitly in what follows.

The average effective action $\Gamma_k$ is obtained from $\Gamma$ by
introducing some kind of infrared cutoff at momentum $k$.
For simplicity one can think at first of a sharp cutoff.
There are various ways of dealing with the determinant in the
denominator, leading to different definitions of $\Gamma_k$.

For example, if we do not introduce the IR cutoff in the denominator
and choose $\phi_0$ to be the minimum of the effective
potential $V_0$ at scale $k=0$, we get an effective potential $V_k$
which satisfies $V_0(\phi_0)=0$ and whose value at the minimum is a
decreasing function of $k$.
(Basically this is because one is removing the zero
point energy of the modes with $|q|<k$.)
The effective potential $V_k(\phi)$ defined in this way
cannot be interpreted as the energy density of a translationally
invariant vacuum. Instead, it represents the energy density
of the system when it is enclosed in a box of size $L=2\pi/k$.
The modes with momentum less than $k$ are removed from the
spectrum, so that for very large $k$ the energy density decreases like
$k^4$. This is the well known Casimir effect [5].
This definition of the effective action $\Gamma_k$ is therefore
relevant in the description of inhomogeneous processes,
for example bubble nucleation.

In this paper we will be interested in another definition.
We will put the IR cutoff also in the denominator and interpret
$\phi_0$ as the minimum of the average potential $V_k$
(from here on we will use the notation $\phi_k$ instead of
$\phi_0$, reserving the notation $\phi_0$ for the minimum of
the usual effective potential $V=V_0$).
With this definition, the minimum of the average potential is
$V_k(\phi_k)=0$, independently of $k$.
Physically, one can think that the system has been enclosed in a
finite box of side $L>2\pi/k$ and
the limit $L\to\infty$ has been taken, with $k$ fixed.
This definition is relevant if we are interested in phenomena
which occur in spacetime regions of size $\sim k^{-1}$, but
the system is not physically confined therein.

Note that the result of the two procedures differs only in the value
of the potential at the minimum. Therefore, it is only in the presence
of gravity that the distinction becomes important.
We also note that with our definition of the average action,
the one loop renormalization group equation becomes exact [6].
Even though we will not write down the exact renormalization group
in what follows, we will keep in mind that with our definition of
the average action the renormalization group equation has a validity
that goes beyond the one loop approximation.

The use of a sharp cutoff has certain disadvantages, so we
will follow [3] and define the average effective action in
the following way. Consider the function
$$
P_k(q^2)={q^2\over 1-f_k^2(q^2)}\ , \eqno(2.3)
$$
where $f_k(q^2)=\exp\left(-a\left({q^2\over k^2}\right)^b\right)$,
for some constants $a$, $b$. The function $P_k(q^2)$ approaches
exponentially the function $q^2$ for $|q|>k$, but tends to $k^2$
(when $b=1$) or diverges (when $b>1$) for $q^2\to 0$.
The average effective action $\Gamma_k$ is obtained from the
ordinary effective action $\Gamma$ by replacing in the loop integrals
the momentum variable $q^\mu$ by $\sqrt{P_k(q^2)}\hat q^\mu$,
where $\hat q^\mu=q^\mu/|q|$.
The effect is that the propagation of the modes with momenta smaller
than $k$ is suppressed.
In the limit $b\to\infty$ the function $f_k$ becomes a step function
and the modes with $|q|<k$ do not propagate at all.
\footnote{$^\dagger$}
{If we were to replace $q^2$ by $P_k(q^2)$ in the numerator of (2.2),
leaving the denominator
alone, one would have $V_0(\phi_0)=0$ for $k=0$, but the minimum of
the potential would be an {\it increasing} function
of $k$, because the contribution of the modes with $|q|<k$ is enhanced.
The physical meaning of this procedure is not clear.}
Our definition of the effective action is then
$$
\Gamma_k={1\over2}\ln
\left({{\rm Det}_k\co(\phi)\over{\rm Det}_k\co(\phi_k)}\right)\ ,
\eqno(2.4)
$$
where $\phi_k$ is the minimum of $V_k$ and ${\rm Det}_k$ is the
determinant with the momentum integration modified as described
above.
With this definition the minimum of the potential is zero
for all scales:
$$
V_k(\phi_k)=0\ .\eqno(2.5)
$$
On the other hand, the derivatives of
$V_k$ with respect to $\phi$ are not affected by the denominator
and therefore all the results which were previously obtained
for scalar theories (in the absence of gravity) [3] remain valid.
For a related discussion of the role of the minimum of the potential
in the renormalization group, see [7].

Einstein's theory is a gauge theory and the definition of the
average effective action involves some complications which are not
present in the pure scalar case, since an IR cutoff will generally
break diffeomorphism invariance.
In the case of gauge theories, this point has been discussed
in [8,9]. We will not discuss these issues here; we shall
follow the approach of [9], where it is shown that
using the background field method one can preserve gauge invariance
with respect to background gauge transformations.

\bigskip
\goodbreak
\leftline{\bf 3. Flow equations}
\smallskip
\noindent
It is obviously impossible to follow the renormalization group flow
of all the parameters appearing in the effective action $\Gamma_k$,
so we will concentrate our attention on the first few terms in the
Taylor expansion of $V_k$ and $Z_k$.
We parametrize the action by the position of the minimum of the
potential $\phi_k$, the quartic coupling at the minimum $\lambda_k$,
and the wave function renormalization at the minimum $\wfr_k$:
$$
V_k'(\phi_k)=0\ \ ,\qquad
\lambda_k=V_k''(\phi_k)\ \ ,\qquad
\wfr_k=Z_k(\phi_k)\ ,\eqno(3.1)
$$
where a prime denotes derivative with respect to $\phi^2$.
The average action (2.1) is thus approximated by assuming
$$
V_k(\phi)={1\over 2}\lambda_k(\phi^2-\phi_k^2)^2\ \ ,
\qquad
Z_k(\phi)=\wfr_k\ \eqno(3.2)
$$
and neglecting all other terms.
We will follow the flow of the parameters $\phi_k$, $\lambda_k$
and $\wfr_k$.

The renormalization group describes the change in the effective
action as one integrates away fluctuations of the fields with
decreasing momentum.
The effective action at the scale $k$ is used as classical action
in the functional integral giving the effective action at a lower scale.
We will therefore assume that the classical action entering in the
definition of the path integral has the form (2.1),
with classical potential
$V(\phi)={1\over 2}\lambda(\phi^2-\phimin^2)^2$
and $Z(\phi)=\wfr$, a constant.
We ignore terms with higher derivatives of the fields, and
higher powers of $\phi$ in $V$ and in $Z$.
After taking the derivative with respect to $k$ we replace the
classical parameters $\phimin$, $\lambda$ and $\wfr$
by their running counterparts $\phi_k$, $\lambda_k$ and $\wfr_k$.
This is the ``renormalization group improvement''.

By taking the derivative of (3.1) with respect to $k$
we get
$$
\eqalignno{
k{\partial\phi^{\, 2}_k\over\partial k}=&
-{1\over\lambda_k}\left(k{\partial V'_k\over\partial
k}\right)\bigg|_{\phi=\phi_k}\equiv k^2\gamma(k)
\ ,&(3.3a)\cr
k{\partial\lambda_k\over\partial k}=&
\ \left(k{\partial V''_k\over\partial k}\right)\bigg|_{\phi=\phi_k}
\phantom{-\lambda_k}\ \equiv\beta (k)
\ ,&(3.3b)\cr
k{\partial\ln\wfr_k\over\partial k}=&
\ \ {k\over\wfr_k}
\left({\partial Z_k\over\partial k}\right)\bigg|_{\phi=\phi_k}
\ \ \ \equiv\eta(k)\ .&(3.3c)\cr}
$$
In (3.3b) and (3.3c) we are neglecting terms
$V_k'''(\phi_k)k{\partial\phi_k^2\over\partial k}$ and
$Z_k'(\phi_k)k{\partial\phi_k^2\over\partial k}$,
which take into account the variation in the point of definition
of $\lambda_k$ and $\wfr_k$.
This is in accordance with the approximations (3.2).

In order to obtain explicit expressions for the beta functions
$k^2\gamma$ and $\beta$
and the anomalous dimension $\eta$, we have to write first the
expressions for $V_k$ and $Z_k$.
If we expand the action around a classical field $\phicl$,
the small fluctuation operator has the form
$$
\co(\phicl)=-\wfr\partial^2+6\lambda\phicl^2-2\lambda\phimin^2\ .\eqno(3.4)
$$
In order to calculate the effective potential one chooses
a constant classical field $\phicl=\bphi$; then
$$
V_k(\bphi)={1\over\Omega}\Gamma_k(\bphi)=
{1\over 2}\int\limits_{|q|^2=0}^{\ \ \Lambda}
{d^4q\over(2\pi)^4}
\ln\left({\wfr P_k(q^2)+6\lambda\bar\phi^2-2\lambda\phimin^2
\over \wfr P_k(q^2)+4\lambda\phimin^2}\right)\ ,
\eqno(3.5)
$$
where $\Omega$ is the spacetime volume.
The beta functions which are defined in the r.h.s. of (3.3a,b)
can be obtained by deriving (3.5) and then
replacing the classical parameters $\phimin$, $\lambda$ and $\wfr$
by their running counterparts $\phi_k$, $\lambda_k$ and $\wfr_k$:
$$
\eqalignno{
\gamma(k)=&\ {1\over32\pi^2}{1\over k^2}\int dx x
{6 \wfr_k \over (\wfr_k P_k+4\lambda_k\phi_k^2)^2}
k{\partial P_k(x)\over\partial k}
={3\over16\pi^2}{1\over\wfr_k}I_{-2}
\left({4\lambda_k\phi_k^2\over\wfr_k}\right)
\ ,&(3.6a)\cr
\beta (k)=&\ {1\over 32\pi^2}\,\int dx x {72\wfr_k\lambda_k^2\over
(\wfr_k P_k(x)+4\lambda_k\phi_k^2)^3}
k{\partial P_k\over\partial k}
={9\over4\pi^2}{\lambda_k^2\over\wfr_k^2}
I_{-3}\left({4\lambda_k\phi_k^2\over\wfr_k}\right)
\ ,&(3.6b)}
$$
where $x=|q|^2$, $P_k=P_k(x)$ and
$$
k^{2(n+3)}I_n(w)=\int dx x(P_k+w)^n k{\partial P_k\over\partial k}\ .\eqno(3.7)
$$
These integrals are related to the integrals $L^d_n(w)$ used in
[3,9] by $I_n(w)=L^4_{-n-1}(w)/(n+1)$.
Since $k{\partial P_k\over\partial k}$ is peaked at $x\approx k^2$
and goes to zero exponentially for $x\to\infty$ and as a power
for $x\to 0$, these integrals are automatically UV and IR convergent.

Let us now derive the expression for the anomalous dimension $\eta$.
The wave function renormalization can be obtained by computing the
effective action for a nonconstant background. We choose
$$
\phicl(x)=\bphi+\epsilon\cos(p\cdot x)\ ,\eqno(3.8)
$$
with $\bphi$ constant.
The wave function renormalization constant is
$$
Z_k(\bphi)=\lim\limits_{p\to 0}\lim\limits_{\epsilon\to 0}
{2\over\Omega}{\partial\over \partial p^2}
{\partial^2\over\partial\epsilon^2}
\Gamma_k(\phicl)
\ .\eqno(3.9)
$$
The small fluctuation operator (3.4) is represented
in momentum space by the kernel
$$
M(q,q')=M_0(q,q')+\epsilon M_1(q,q')+\epsilon^2 M_2(q,q')\ ,\eqno(3.10)
$$
where
$$
\eqalign{
M_0(q,q')=&\,\left(\wfr q'^2+6\lambda\bphi^2
-2\lambda\phimin^2\right)\delta(q+q')\ ,\cr
M_1(q,q')=&\,6\lambda\bphi\left(\delta(q+q'+p)+\delta(q+q'-p)\right)\ ,\cr
M_2(q,q')=&\,3\lambda\delta(q+q')
+{3\over2}\lambda\left(\delta(q+q'+2p)+\delta(q+q'-2p)\right)\ .\cr}
\eqno(3.11)
$$
We have
$$
\ln{\rm Det}\co=\tr\ln M=\tr\ln M_0+\epsilon\,\tr\, M_0^{-1}M_1
+\epsilon^2\left(\tr\, M_0^{-1}M_2
-{1\over2}\tr\, M_0^{-1}M_1\,M_0^{-1}M_1\right)
+\ldots
\eqno(3.12)
$$
where $\tr$ denotes the functional trace.
The determinants appearing in (2.4) are obtained by replacing
$\tr$ with $\tr_k$, a functional trace in which the
momentum integrations are modified as described in the previous
section.
The first term in (3.12) then reproduces the potential (3.5).
The term of order $\epsilon$ and the first term
of order $\epsilon^2$ are zero.
The remaning term gives, after some manipulations,
$$
\eqalign{
Z_k(\bphi)=&-72\lambda^2\bphi^2\lim\limits_{p\to 0}
{\partial\over\partial p^2}
\int {d^4q\over(2\pi)^4}
{1\over (\wfr P_k(q^2)+6\lambda\bphi^2-2\lambda\phimin^2)
(\wfr (\sqrt{P_k}\hat q+p)^2+6\lambda\bphi^2-2\lambda\phimin^2)}
\cr=&
\ 144\lambda^3\wfr\bphi^2(3\bphi^2-\phimin^2)
\int {d^4q\over(2\pi)^4}
{1\over(\wfr P_k(q^2)+6\lambda\bphi^2-2\lambda\phimin^2)^4}\ .}
\eqno(3.13)
$$
We have used the fact that upon symmetric integration,
the integral in the first line reduces to a function of
$p^2$ only. One can then use ${\partial\over\partial p^2}=
{1\over8}{\partial\over\partial p_\lambda}{\partial\over\partial p^\lambda}$.
This definition of $Z_k$ differs from the one used in [3],
where $P_k((q+p)^2)$ is used instead of $(\sqrt{P_k}\hat q+p)^2$.
It simplifies the calculations considerably and leads practically
to the same final results. This simplification will be important
in Section 5.

According to (3.3c), the anomalous dimension is obtained by deriving
(3.13) with respect to $k$, replacing $\phimin$, $\lambda$
and $\wfr$ by their running counterparts $\phi_k$, $\lambda_k$
and $\wfr_k$, and setting also $\bphi=\phi_k$:
$$
\eta(k)=
-{72\lambda_k^3\wfr_k\phi_k^4\over\pi^2}
\int dx x{1\over(\wfr_k P_k(x)+4\lambda_k\phi_k^2)^5}
k{\partial P_k\over\partial k}
=-{72\lambda_k^3\phi_k^4\over\pi^2\wfr_k^4 k^4}
I_{-5}\left({4\lambda_k\phi_k^2\over\wfr_k}\right)
\ .\eqno(3.14)
$$
It is convenient to use the rescaled field variables
$\tphi(x)=\sqrt{\wfr_k}\phi(x)$ and the rescaled coupling constant
$\tlambda_k=\wfr_k^{-2}\lambda_k$.
The equations (3.3) can be rewritten
$$
\eqalignno{
k{\partial\tphi^{\, 2}_k\over\partial k}=&
\ \eta\tphi_k^2+{3\over16\pi^2}k^2I_{-2}
\left(4\tlambda_k\tphi_k^2\right)
\ ,&(3.15a)\cr
k{\partial\tlambda_k\over\partial k}=&
-2\eta\tlambda_k+{9\over4\pi^2}\tlambda_k^2
I_{-3}\left(4\tlambda_k\tphi_k^2\right)
\ ,&(3.15b)\cr
k{\partial\ln \wfr_k\over\partial k}=&
\ -{72\tlambda^3\tphi_k^4\over\pi^2 k^4}
I_{-5}\left(4\tlambda_k\tphi_k^2\right)
\ .&(3.15c)\cr}
$$
Note that $\wfr_k$ does not appear in the r.h.s. anymore.

It is not possible to give a solution of the resulting
system of p.d.e.'s in closed form. However, analytic solutions can be
obtained for $k^2$ very large or very small.
Let us consider first the anomalous dimension $\eta$.
Since the function $k{\partial P_k\over\partial k}$
is peaked at $q\approx k$ and goes to zero exponentially for large
$q$, and as a power for small $q$,
the main contribution to the integrals (3.6) and (3.14)
comes from the region $q\approx k$, where $P_k(q^2)\approx k^2$.
In this region, for $k^2\gg\phi_k^2$, we can neglect
$\phi_k^2$ with respect to $P_k$. One can therefore approximately
evaluate $I_{-5}$ at $\phi_k=0$, so the integral gives only a numerical
coefficient. We see that
$\eta=-{72\tlambda_k^3I_{-5}(0)\over\pi^2}{\tphi_k^4\over k^4}\ll1$.
In the case $k^2\ll\phi_k^2$, by a similar reasoning, we can neglect
the term $P_k$ with respect to $\phi_k$ in the denominator.
In this regime
$\eta=-{9I_0(0)\over 128\pi^2}{k^6\over\tlambda_k^2\tphi^6}\ll1$.
So in both regimes the anomalous dimension $\eta$ is small.
This is in accordance with the analysis of the exact RG given
in [6].

Let us now consider the equations (3.15a,b).
Assume again that for large $k^2$ the mass terms in the
denominators can be neglected with respect to factors $P_k$.
Neglecting also the anomalous dimension,
the equations (3.15) then reduce to the following:
$$\eqalignno{
k{\partial\tphi_k^2\over\partial k}=&
{3I_{-2}(0)\over16\pi^2}k^2
\ ,&(3.16a)\cr
k{\partial\tlambda_k\over\partial k}=&
{9I_{-3}(0)\over4\pi^2}\tlambda_k^2
\ ,&(3.16b)}
$$
Using that $I_{-3}(0)=1$, independently of $a$ and $b$,
(3.16) gives the usual logarithmic running of the quartic coupling at high
energies, whereas $\phi_k$ scales like $k$, as dimensional arguments
would suggest.

This last result seems to invalidate the approximation $k^2\gg\phi_k^2$
and to cast doubts on the consistency of these results.
However, if we write $\phi_k^2=c\ k^2$, with $c$ a constant,
and insert in (3.6,14), it is easy to see that the conclusions
of the previous analysis are confirmed; only the numerical coefficients
appearing in (3.16) would be modified.

Le us now consider the opposite limit: $k^2\ll\tphi_k^2$.
This is the limit $k\to 0$, when $\tphi_0\not=0$.
The beta functions become
$\gamma(k)={3I_0\over256\pi^2}{1\over\tlambda_k^2}{k^4\over\tphi_k^4}\ll1$
and
$\beta(k)={9I_0\over256\pi^2}{1\over\tlambda_k}{k^6\over\tphi_k^6}\ll1$.
In equations (3.15a,b) the anomalous dimension terms are of the same
order as the other terms. We get
$$
\eqalignno{
k{\partial\tphi_k^{\, 2}\over\partial k}=&
-{15I_0(0)\over256\pi^2}{1\over\tlambda_k^2}{k^6\over\tphi_k^4}
\ ,&(3.17a)\cr
k{\partial\tlambda_k\over\partial k}=&
{45I_0(0)\over256\pi^2}{1\over\tlambda_k}{k^6\over\tphi_k^6}
\ .&(3.17b)}
$$
The running of $\lambda_k$ and $\tphi_k^2$ is damped by
powers of $k^2/\tphi_k^2$ and stops for $k\to 0$.
The solutions for small $k$ are
$$
\eqalignno{
\tphi_k^2=&\,\tphi_0^2\left[
1-{5I_0\over512\pi^2}{1\over\tlambda_0^2}
{k^6\over\tphi_0^6}\right]\ ,&(3.18a)\cr
\tlambda_k=&\,\tlambda_0\left[
1+{15I_0\over512\pi^2}{1\over\tlambda_0^2}
{k^6\over\tphi_0^6}\right]\ .&(3.18b)\cr}
$$
\bigskip
\goodbreak
\leftline{\bf 4. The effect of gravity on the beta functions}
\smallskip
\noindent
Let us now turn the gravitational field on.
We add to the action the Einstein-Hilbert term, so the
total classical action becomes
$$
S(\phi,g)=
\int d^4x\,\sqrt{g}\,
\left[{1\over2}\wfr g^{\mu\nu}\partial_\mu\phi\partial_\nu\phi+V(\phi)
+\kappa R\right]
\ ,\eqno(4.1)
$$
where $\kappa=(1/16\pi G)$ and
$V(\phi)={1\over2}\lambda(\phi^2-\phimin^2)$.
We treat the metric as a quantum field, but we will not take
into account the running of Newton's constant.

We expand the fields $\phi$ and $g_{\mu\nu}$:
$$
\phi=\phicl+\delta\phi\ \ ;\qquad
g_{\mu\nu}=\delta_{\mu\nu}+{1\over\sqrt\kappa} h_{\mu\nu}\ .
\eqno(4.2)
$$
The linearized Euclidean action is a quadratic form
which can be written, after Fourier transforming
(we use $\partial_\mu\to iq_\mu$)
$$
S^{(2)}(\psi,\omega,\sigma;\phi)
=\ {1\over2}\int d^4q\int d^4 q' \sum\limits_{A,B}
\ \Phi_A(q)\cdot{\cal O}_{[AB]}\cdot\Phi_B(q')\ ,\eqno(4.3)
$$
where the indices $A$, $B$ label the two types of fields
$\Phi_1=h_{\mu\nu}$, $\Phi_2=\delta\phi$
and the dots stand for contraction over the tensor indices.
When written out explicitly in terms of the components
of the fields, ${\cal O}$ is a $11\times 11$ matrix.
When $V=0$, this linearized action is invariant
under linearized gauge transformations.
Let $x'^\mu=x^\mu-v^\mu$ be an infinitesimal
coordinate transformation. The variations of the fields are
$$
\delta g_{\mu\nu}=\partial_\mu v_\nu+\partial_\nu v_\mu\ ,\qquad
\delta\phi=0\ . \eqno(4.4)
$$
There follows that the fields
$$
h_{\mu\nu}=i(q_\mu v_\nu+q_\nu v_\mu)\ ,\qquad
\delta\phi=0 \eqno(4.5)
$$
are null vectors for the operator ${\cal O}$.
We choose the following gauge-fixing term:
$$
S_{\rm GF}=
{1\over2\alpha}\int d^4x\,
\partial_\mu h^{\mu\nu}\partial^\sigma h_{\sigma\nu}\ .
\eqno(4.6)
$$
In this gauge the ghost determinant is field indipendent, so
it will be neglected in what follows.

To compute the one-loop effective action one now needs to calculate
the functional determinant of the operator $\cal O$ appearing in the
previous formulas and (3.2).
It is convenient to use the method of the spin projector operators.
Choose a coordinate system such that $x^L$ is in the direction of the
momentum and $x^i$ are transverse coordinates.
In these coordinates the momentum has components $q^\mu=(|q|,0,0,0)$.
One can decompose the tensor $h_{\mu\nu}$ into the fields
$h_{LL}$, $h_{(Li)}$, $h=\sum_k h_{kk}$ and $\tilde h_{ij}=
h_{ij}-{1\over3}\delta_{ij}h$, carrying spin and parity
$0^+$, $1^-$, $0^+$ and $2^+$ respectively. The field $\delta\phi$
obviously has spin-parity $0^+$.
A complete set of spin projectors for this system is listed
in the Appendix.

The total linearized quadratic action, including the gauge-fixing, ghost
and potential terms, can be rewritten as
$$
S^{(2)}={1\over 2}\int d^4q\ \Phi_A(-q)\cdot
a_{ij}^{AB}(J^{\cal P})\, P_{ij}^{AB}(J^{\cal P})\cdot \Phi_B(q)\ ,
\eqno(4.7)
$$
where $a_{ij}^{AB}(J^{\cal P})$ are coefficient matrices,
representing the inverse propagators of each set of fields with
definite spin and parity. They are
$$
\eqalignno{
a(2^+)=&-{1\over2}\left(q^2+{V\over\kappa}\right)\ ,&(4.8a)\cr
&\null\cr
a(1^-)=&\,{1\over2}\left({1\over\alpha}q^2-{V\over\kappa}\right)
\ ,&(4.8b)\cr
&\null\cr
a(0^+)=&\ \left[
\matrix{& q^2+{1\over4\kappa}V
        & {\sqrt3\over4\kappa}V
        & \sqrt{3\over\kappa}\phi V'
    \cr
        & {\sqrt3\over4\kappa}V
        & {1\over\alpha}q^2-{1\over4\kappa}V
        & {1\over\sqrt\kappa}\phi V'
    \cr
        & \sqrt{3\over\kappa} \phi V'
        & {1\over\sqrt\kappa}\phi V'
        & \wfr q^2+2 V'+4\phi^2 V''
                               \cr}   \right]\ . &(4.8c)\cr}
$$
The matrix elements of $a(0^+)$ refer to the fields
$h_{LL}$, $h$ and $\delta\phi$, in this order.
Taking into account the multiplicity of these contributions,
the one-loop effective potential is now
$$
V_k(\phi)={1\over2}\int {d^4q\over(2\pi)^4}
\left[5\ln\left({a_k(2^+)(\phi)\over a_k(2^+)(\phi_k)}\right)
+3\ln\left({a_k(1^-)(\phi)\over a_k(1^-)(\phi_k)}\right)
+\ln\left({\det a_k(0^+)(\phi)\over \det a_k(0^+)(\phi_k)}\right)\right]
\ ,\eqno(4.9)
$$
where the modified inverse propagators $a_k$ are obtained from
the $a_k$'s given in (4.8) by replacing $q^2$ with $P_k(q^2)$.

Proceeding as in the previous section we find for the beta functions
$$
\eqalignno{
\gamma(k)=&
{1\over 32\pi^2}{1\over k^2}\int dx x
{6\wfr_k\over (\wfr_k P_k(x)+4\lambda_k\phi_k^2)^2}
k{\partial P_k\over \partial k}
\ ,&(4.10a)\cr
\beta (k)=&
{1\over 32\pi^2}\int dx x
{1\over 4\kappa P_k(x)^2(\wfr_k P_k(x)+4\lambda_k\phi_k^2)^3}
\biggl[(13\alpha-21)\wfr_k^3\lambda_k P_k(x)^3\cr
&
+4\wfr_k\lambda_k^2(72\kappa+(43\alpha-51)\wfr_k\phi_k^2)P_k(x)^2
+720(\alpha-1)\wfr_k\lambda_k^3\phi_k^4 P_k(x)
+960(\alpha-1)\lambda_k^4\phi_k^6\biggr]
k{\partial P_k\over \partial k}\qquad
\,.&(4.10b)}
$$
Note that (4.10a) is identical to (3.6a) and the term containing
$\kappa$ in the numerator of (4.10b) reproduces the beta function
of the pure scalar theory, (3.6b).

\bigskip
\leftline{\bf 5. The effect of gravity on the wave function renormalization}
\smallskip
\noindent
We have to compute the wave function renormalization constant $\wfr_k$
in the presence of gravitons. It is given again by (3.9), with the
effective action now including the effect of graviton loops.
The calculation begins with the expansion of the classical action
(4.1) around the background (4.2), with $\phicl$ now given by (3.8).
The linearized action reads
$$
\eqalign{
S^{(2)}={1\over2}\int dx\Biggl\{ &
h_{\mu\nu}\biggl[
\left(\delta^{\mu\nu}\partial^\rho\partial^\sigma
-{1\over2}\delta^{\mu\nu}\delta^{\rho\sigma}\partial^2
-\delta^{\nu\sigma}\partial^\mu\partial^\rho
+{1\over2}\delta^{\mu\rho}\delta^{\nu\sigma}\partial^2\right)\cr
&
+\left({\wfr\over2\kappa}(\partial\phicl)^2+{V\over\kappa}\right)
\left({1\over4}\delta^{\mu\nu}\delta^{\rho\sigma}
-{1\over2}\delta^{\mu\rho}\delta^{\nu\sigma}\right)
-{\wfr\over2\kappa}\delta^{\mu\nu}\partial^\rho\phicl\partial^\sigma\phicl
+{\wfr\over\kappa} \delta^{\mu\rho}\partial^\nu\phicl\partial^\sigma\phicl
\biggr]h_{\rho\sigma}
\cr &
+h_{\mu\nu}{1\over\sqrt\kappa}\biggl[
\wfr \delta^{\mu\nu}\partial^\lambda\phicl\partial_\lambda
-2\wfr\partial^\mu\phicl\partial^\nu
+\delta^{\mu\nu}{dV\over d\phi}\biggr|_{\phicl}\biggr]\delta\phi
\cr&
\qquad\qquad\qquad
+\delta\phi\biggl[-\wfr\partial^2+{d^2 V\over d\phi}\biggr|_{\phicl}
\biggr]\delta\phi
\Biggr\}
}\eqno(5.1)
$$
Next one has to use the Taylor expansion of $V$, ${dV\over d\phi}$
and ${d^2V\over d\phi^2}$ around $\bar\phi$, keeping terms
up to order $\epsilon^2$ (this involves
derivatives of $V$ up to fourth order).
Upon Fourier transforming, the result can be recast in the form (4.3).
The operator $\co$ contains again terms up to second order in
$\epsilon$, and can be written as in (3.10), where $M_0$, $M_1$
and $M_2$ are now matrices in the space of the
fields $h_{\mu\nu}$ and $\delta\phi$.
The coefficients $M$ are conveniently displayed as
matrices of the form
$$
M=\left[\matrix{M_{\mu\nu}{}^{\rho\sigma}
&M_{\mu\nu}{}^{\cdot}\cr
M_{\cdot}{}^{\rho\sigma}
&M_{\cdot}{}^{\cdot}\cr}\right]\ ,\eqno(5.2)
$$
where the dots label the entries that correspond to $\delta\phi$.
It is convenient to write
$M_0(q,q')=\delta(q+q')\hat M_0(q')$,
with
$$
\hat M_0=a(2^+)P(2^+)+a(1^-)P(1^-)
+a_{ij}(0^+)P_{ij}(0^+)\ ,\eqno(5.3)
$$
where $i,j=1,2,3$ refer to $h_{LL}$, $h$ and $\delta\phi$.
The matrix structure (5.2) is carried by the projectors.
Not that (5.3) is the kernel which appears in (4.7).
The advantage of this way of writing is that
$M_0^{-1}(q',q'')=\delta(q'+q'')\hat M_0(q')^{-1}$,
where
$$
\hat M_0^{-1}=a(2^+)^{-1}P(2^+)+a(1^-)^{-1}P(1^-)
+a_{ij}(0^+)^{-1}P_{ij}(0^+)\ ,\eqno(5.4)
$$
$a^{-1}$ being the inverses of the matrices given in (4.8).
We can write
$$
M_1(q,q')=M_1^+(q')\delta(q+q'+p)+
M_1^-(q')\delta(q+q'-p)\ ,\eqno(5.5)
$$
where
$$
\eqalignno{
M_1^+(q')_{\mu\nu}{}^{\rho\sigma}=&\,
{1\over2\kappa}{dV\over d\phi}
\left({1\over4}\delta_{\mu\nu}\delta^{\rho\sigma}
-{1\over2}\delta_\mu^\rho\delta_\nu^\sigma\right)\ ,&(5.6a)\cr
M_1^+(q')_{\mu\nu}{}^{\cdot}=&\,
{1\over2\sqrt\kappa}\left(\wfr p_{(\mu} q'_{\nu)}
-{1\over2}\wfr \delta_{\mu\nu}p\cdot q'
+{1\over2}{d^2V\over d\phi^2}\delta_{\mu\nu}\right)\ ,&(5.6b)\cr
M_1^+(q')_{\cdot}{}^{\rho\sigma}=&\,
{1\over2\sqrt\kappa}\left(-\wfr p^\rho p^\sigma
+{1\over2}\wfr \delta^{\rho\sigma}p^2
-\wfr p^{(\rho} q'^{\sigma)}
+{1\over2}\delta^{\rho\sigma}p\cdot q'
+{1\over2}{d^2V\over d\phi^2}\delta^{\rho\sigma}\right)\ ,&(5.6c)\cr
M_1^+(q')_\cdot{}^\cdot=&\,{1\over2}{d^3V\over d\phi^3}\ ,&(5.6d)\cr}
$$
and
$$
\eqalignno{
M_1^-(q')_{\mu\nu}{}^{\rho\sigma}=&\,
{1\over2\kappa}{dV\over d\phi}
\left({1\over4}\delta_{\mu\nu}\delta^{\rho\sigma}
-{1\over2}\delta_\mu^\rho\delta_\nu^\sigma\right)\ ,&(5.7a)\cr
M_1^-(q')_{\mu\nu}{}^{\cdot}=&\,
{1\over2\sqrt\kappa}\left(-\wfr p_{(\mu} q'_{\nu)}
+{1\over2}\wfr \delta_{\mu\nu}p\cdot q'
+{1\over2}{d^2V\over d\phi^2}\delta_{\mu\nu}\right)\ ,&(5.7b)\cr
M_1^-(q')_{\cdot}{}^{\rho\sigma}=&\,
{1\over2\sqrt\kappa}\left(-\wfr p^\rho p^\sigma
+{1\over2}\wfr \delta^{\rho\sigma}p^2
+\wfr p^{(\rho} q'^{\sigma)}
-{1\over2}\delta^{\rho\sigma}p\cdot q'
+{1\over2}{d^2V\over d\phi^2}\delta^{\rho\sigma}\right)\ ,&(5.7c)\cr
M_1^-(q')_\cdot{}^\cdot=&{1\over2}{d^3V\over d\phi^3}\ ,&(5.7d)\cr}
$$
All derivatives of $V$ are evaluated at $\bar\phi$.
Finally
$$
M_2(q,q')=\hat M_2(q')\delta(q+q')+
\tilde M_2(q')\left(\delta(q+q'+2p)+\delta(q+q'-2p)\right)\ ,\eqno(5.8)
$$
where
$$
\eqalignno{
\hat M_2(q')_{\mu\nu}{}^{\rho\sigma}=&\,
{1\over2\kappa}\left(\wfr \delta_\mu^\rho p_\nu p^\sigma
-{1\over2}\wfr\delta_{\mu\nu}p^\rho p^\sigma\right)
+{1\over4}\left(\wfr p^2+{d^2V\over d\phi^2}\right)
\left({1\over4}\delta_{\mu\nu}\delta^{\rho\sigma}
-{1\over2}\delta_\mu^\rho\delta_\nu^\sigma\right)\ ,&(5.9a)\cr
\hat M_2(q')_{\mu\nu}{}^{\cdot}=&\,
{1\over8\sqrt\kappa}{d^3V\over d\phi^3}\delta_{\mu\nu}\ ,&(5.9b)\cr
\hat M_2(q')_{\cdot}{}^{\rho\sigma}=&\,
{1\over8\sqrt\kappa}{d^3V\over d\phi^3}\delta^{\rho\sigma}\ ,&(5.9c)\cr
\hat M_2(q')_\cdot{}^\cdot=&\,{1\over4}{d^4V\over d\phi^4}\ ,&(5.9d)\cr}
$$
As in section 3, $\tr\, M_0^{-1}M_1=0$ and $\tr\, M_0^{-1}M_2=0$.
(The explicit form of $\tilde M_2$ is not needed to obtain the latter
result). The remaining term entering in $Z_k$ is
$$
\tr\, M_0^{-1} M_1\, M_0^{-1} M_1=
2\Omega\int {d^4q\over(2\pi)^4}
{\rm tr}\,\left(M_1^-(q)\hat M_0^{-1}(q)
M_1^+(q-p)\hat M_0^{-1}(q-p)\right)\ ,\eqno(5.10)
$$
where tr denotes the trace over the matrix indices in the sense
of (5.2).
There remains to take the second derivative of this expression
with respect to $p^\mu$ and evaluate at $p^\mu=0$.
This is the most tedious part of the calculation.
It is made slightly easier by considering only $\bar\phi=\phimin$.
Then $V=V'=0$ and the matrix (4.8c) becomes diagonal.
Furthermore in (5.6-7) one can put
${dV\over d\phi}=0$, ${d^2V\over d\phi^2}=4\lambda\phimin^2$,
${d^3V\over d\phi^3}=12\lambda\phimin$, ${d^4V\over d\phi^4}=12\lambda$.
The final result is
$$
\eqalign{
Z_k(\phimin)=&
-{1\over4\kappa}\int {d^4q\over(2\pi)^4}
{1\over\wfr P_k(q^2)+4\lambda\phimin^2}
\Biggl[{(13\alpha+3)\over8}\wfr^2
+(4\alpha+6){\wfr\lambda\phimin^2\over P_k(q^2)}
-4(3+\alpha){\lambda^2\phimin^4\over P_k(q^2)^2} \cr
&\qquad\ \
+{\alpha-3\over2}{\lambda\wfr^2\phimin^2\over\wfr P_k(q^2)+4\lambda\phimin^2}
-{32(3+\alpha)\lambda^3\wfr\phimin^6\over (\wfr P_k(q^2)+4\lambda\phimin^2)^2}
-{1152\lambda^3\wfr\kappa\phimin^4\over(\wfr P_k(q^2)+4\lambda\phimin^2)^3}
\Biggr]\ .\cr}\eqno(5.11)
$$
Note that the last term reproduces the result (3.13) of a pure
scalar theory. The remaining terms are all of order $1/\kappa$.
The corresponding anomalous dimension is
$$
\eqalign{
\eta(k)=&
{1\over64\pi^2\kappa}\int dx x
{1\over8P_k(x)^3(\wfr P_k(x)+4\lambda_k\phi_k^2)^5}
\biggl[(13\alpha+3)\wfr^5 P_k(x)^6
+12(19\alpha+9)\lambda_k\wfr_k^4\phi_k^2 P_k(x)^5\cr
&
\qquad
+48(31\alpha\wfr+21)\lambda_k^2\wfr_k^3\phi_k^4 P_k(x)^4
+64((49\alpha-9)\wfr_k\phi_k^2-576\kappa)\lambda_k^3\wfr_k\phi_k^4
P_k(x)^3 \cr
&
\qquad
-2560(\alpha+9)\lambda_k^4\wfr_k\phi_k^8 P_k(x)^2
-2048(7\alpha+27)\lambda_k^5\phi_k^{10} P_k(x)
-16384(\alpha+3)\lambda^6\wfr_k^{-1} \phi_k^{12}
\biggr]k{\partial P_k\over\partial k} \ .}
\eqno(5.12)
$$
The term containing $\kappa$ in the numerator reproduces
the anomalous dimension of the pure scalar theory, (3.14).
\bigskip
\leftline{\bf 6. Discussion}
\smallskip
\noindent
The renormalization group equation for the average effective
action of the scalar fields, taking into account the graviton
contribution, is given by (3.3), with the beta functions
(4.10) and anomalous dimension (5.12).
Because there are now two mass scales in this problem,
the behaviour is more complicated than in the pure scalar case:
intermediate mass scales appear.

Let us discuss first the anomalous dimension.
There are four relevant mass scales:
$\phi_k^{8/3}\kappa^{-1/3}\ll\phi_k^2
\ll\phi_k^{4/3}\kappa^{1/3}\ll\kappa$,
dividing the energy range from zero to the Planck mass in
four domains.

For $\phi_k^{4/3}\kappa^{1/3}\ll k^2 \ll\kappa$,
the first term in the numerator and in the denominator
of (5.12) dominate.
For $\phi^2\ll k^2 \ll\phi_k^{4/3}\kappa^{1/3}$ the
term containing $\kappa$ in the numerator dominates over
the first term, which is of order $P_k^6\approx k^{12}$.
For $\phi_k^{8/3}\kappa^{-1/3}\ll k^2 \ll\phi_k^2$
the term containing $\kappa$ in the numerator dominates
over the last term, which is of order $\phi_k^{12}$,
while the $\phi_k^2$ term dominates in the denominator.
Finally, for $k^2 \ll\phi_k^{8/3}\kappa^{-1/3}$
the terms with the highest power of $\phi_k$ dominate.
$$
\eqalignno{
\phi_k^{4/3}\kappa^{1/3}\ll k^2 \ll\kappa\qquad\qquad\ \ \ \
&\eta={(13\alpha+3)\over 512\pi^2}I_{-2}(0)
{k^2\over\kappa}\,\ll 1\ ,&(6.1a)\cr
\phi_k^2\ll k^2 \ll\phi_k^{4/3}\kappa^{1/3}\qquad\ \,
&\eta=-{72\tlambda_k^3\over\pi^2}I_{-5}(0)
{\tphi_k^4\over k^4}\,\ll 1\ ,&(6.1b)\cr
\phi_k^{8/3}\kappa^{-1/3}\ll k^2 \ll\phi_k^2\qquad\qquad\ \
&\eta=-{9\over8\tlambda_k^2\pi^2}I_0(0)
{k^6\over\tphi_k^6}\,\ll 1\ ,&(6.1c)\cr
k^2 \ll\phi_k^{8/3}\kappa^{-1/3}\qquad
&\eta=-{(\alpha+3)\tlambda_k\over 2\pi^2}I_{-3}(0)
{\tphi_k^2\over\kappa}\,\ll1\ .&(6.1d)\cr}
$$
In all cases, the anomalous dimension is small.

We have already observed that the function $\gamma$ describing the
running of the v.e.v. of the scalar field
is not modified by the presence of gravity.
The running of the v.e.v.
of the field is given again by (3.16a) for $k^2\gg\phi_k^2$
and by (3.17a) for $k^2\ll\phi_k^2$.

In discussing the function $\beta(k)$ one has to distinguish
three regimes, separated by the scales
$\phi_k^3\kappa^{-1/2}\ll\phi_k^2\ll\kappa$.
For $\phi_k^2\ll k^2 \ll\kappa$ the dominant terms in (4.10b)
are the ones with the highest power of $P_k$.
Expanding (4.10b) to first order in $k^2/\kappa$ we get
$$
\beta(k)={9\lambda_k^2\over4\pi^2\wfr_k^2}
+{(13\alpha-21)\over 128\pi^2}I_{-2}(0)\lambda_k{k^2\over\kappa}\ .
\eqno(6.2)
$$
The second term is of the same order as the anomalous dimension
(6.1a). Putting together in (3.15b) we find
$$
k{\partial\tlambda_k\over\partial k}=
{9\tlambda_k^2\over4\pi^2}
+{(13\alpha-45)\over 256\pi^2}
I_{-2}(0)\tlambda_k{k^2\over\kappa}\ .
\eqno(6.3)
$$
This seems to show that when the energy approaches the Planck
energy, the coupling begins to run much faster than logarithmically.
This conclusion should be taken with some care, however, since at
energies comparable to Planck's energy the validity of Einstein's
theory as an effective theory becomes questionable.
We expect the coupling constant to run again logarithmically
above the Planck scale, but with a different coefficient that
will depend on the details of the ``new physics'' that one encounters
in this regime. The power--like behaviour indicated by (6.3)
is probably limited to the threshold region.

For $\phi_k^3\kappa^{-1/2}\ll k^2 \ll\phi_k^2$ the term containing
$\kappa$ in the numerator is the dominant one. We find
$$
\beta(k)={9\wfr_k\over256\pi^2}I_0(0){1\over\lambda_k}
{k^6\over\phi_k^6}\ll 1\ .
\eqno(6.4)
$$
Just below the mass threshold of the scalar particles, at
$k^2=\phi_0^2$, this
is of the same order as the anomalous dimension (6.1c) and
$$
k{\partial\tlambda_k\over\partial k}
={785\over256\pi^2}I_0(0){1\over\tlambda_k}
{k^6\over\tphi_k^6}\ .
\eqno(6.5)
$$
Finally, for $k^2\ll\phi_k^3\kappa^{-1/2}$ the terms with the highest
power of $\phi_k$ dominate and
$$
\beta(k)={15(\alpha-1)\over128\pi^2}
I_{-2}(0)\lambda_k{k^2\over\kappa} \ll 1\ .
\eqno(6.6)
$$
This is much smaller than the anomalous dimension (6.1d),
so in the extreme infrared the running of the coupling
constant is dominated by $\eta$. Since $\tphi_k\approx\tphi_0$,
$$
k{\partial\tlambda_k\over\partial k}
={2(\alpha+3)\over 2\pi^2}{\tphi_0^2\over\kappa}\tlambda_k^2\ll1\ .
\eqno(6.7)
$$
The solution has the form
$$
\tlambda_k^2={\tlambda_0^2\over
1-{2(\alpha+3)\over2\pi^2}{\tphi_0^2\over\kappa}\tlambda_0^2
\ln\left(k\over k_0\right)}\ ,
\eqno(6.8)
$$
and therefore $\tlambda_k$ tends (very slowly)
to zero as $k$ goes to zero.
This should be contrasted with equation (3.18b) of the
pure scalar theory, where $\tlambda_k$ tends to a constant.
The different behaviour is due to the presence of a massless
particle, the graviton.

To summarize our results, we have found that the v.e.v.
runs quadratically and the coupling constant runs
logarithmically above the mass of the scalar particles,
while the running is suppressed at energies much lower than
the mass of the scalar.
The details of the residual running at low energies
seem to differ above and below a certain mass scale
$\phi_k^{3/2}\kappa^{-1/4}$. If the scalar has a mass in
the electroweak range, this scale is of the order of
10MeV, while if the scalar has a mass in the GUT range,
this scale is of the order of 10$^{10}$GeV.
While probably of little practical significance, the
appearance of this additional scale is theoretically
quite intriguing.

\bigskip
\centerline{\bf Acknowledgements}
We would like to thank Gabriele Veneziano for the kind
hospitality and support at CERN, where this work was begun.
\bigskip
\goodbreak

\leftline{\bf Appendix: Spin-projector operators}
\medskip
\noindent
For completeness, we list here the explicit expressions
of the spin-projector operators $P_{ij}^{AB}(J^{\cal P})$
that appear in (4.7):
$$ \eqalign{
& P^{hh}(2^+)_{\rho\sigma}{}^{\alpha\beta}=
  T_{(\rho}^{(\alpha}T_{\sigma)}^{\beta)}-
  {1\over3}T_{\rho\sigma}T^{\alpha\beta}\,\ ,\cr
& P^{hh}(1^-)_{\rho\sigma}{}^{\alpha\beta}=
  2\, \t_{(\rho}^{(\alpha}\,\l_{\sigma)}^{\beta)}\ ,\cr
& P^{hh}_{11}(0^+)_{\rho\sigma}{}^{\alpha\beta}=
 {1\over3}\,\t_{\rho\sigma}\,\t^{\alpha\beta}\ ,\cr
& P^{hh}_{12}(0^+)_{\rho\sigma}{}^{\alpha\beta}=
 {1\over\sqrt{3}}\, \t_{\rho\sigma}\,\l^{\alpha\beta}\ ,\cr
& P^{h \phi}_{13}(0^+)_{\rho\sigma}{}^{\cdot}=
 {1\over\sqrt{3}}\, \t_{\rho\sigma}\,\cr
& P^{hh}_{21}(0^+)_{\rho\sigma}{}^{\alpha\beta}=
 {1\over\sqrt{3}}\, \l_{\rho\sigma}\, \t^{\alpha\beta}\ ,\cr
& P^{hh}_{22}(0^+)_{\rho\sigma}{}^{\alpha\beta}=
 \l_{\phi h}\, \l^{\alpha\beta}\ ,\cr
& P^{h \phi}_{23}(0^+)_{\rho\sigma}=
 \l_{\rho\sigma}\ ,\cr
& P^{\phi h}_{31}(0^+)^{\alpha\beta}=
 {1\over\sqrt{3}}\, \t^{\alpha\beta}\ ,\cr
& P^{\phi h}_{32}(0^+)^{\alpha\beta}=
 \l^{\alpha\beta}\ ,\cr
& P^{\phi \phi}_{33}(0^+)=1 \ ,\cr}
$$
where
$$
{\hat q}^\mu=q^\mu/\sqrt{q^2}\ ,\qquad \l_\mu^\nu={\hat q}_\mu{\hat q}^\nu\ ,
\qquad \t_\mu^\nu=\delta_\mu^\nu-\l_\mu^\nu\ .
$$
\bigskip
\goodbreak

\centerline{\bf References}
\bigskip
\noindent
\item{1.} K.G. Wilson and J.B. Kogut, Phys. Rep. {\bf 12C}, 75
(1974);\hfil\break
K.G. Wilson, Rev. Mod. Phys. {\bf 47}, 774 (1975).
\smallskip
\item{2.} J. Polchinski, Nucl. Phys. {\bf B 231}, 269 (1984).
\smallskip
\item{3.} A. Ringwald and C. Wetterich,
Nucl. Phys. {\bf B 334}, 506 (1990);
\hfil\break
C. Wetterich, Nucl. Phys. {\bf B 352}, 529 (1991);
Z. Phys. C {\bf 57}, 451 (1993); {\it ibid.} {\bf C 60}, 461 (1993);
\hfil\break
N. Tetradis and C. Wetterich, Nucl. Phys. {\bf B 422},
541 (1994).
\smallskip
\item{4.} R. Floreanini and R. Percacci,
Nucl. Phys. {\bf B 436}, 141 (1995);
``The renormalization group flow of the dilaton potential'',
to appear in Phys. Rev. D.
\smallskip
\item{5.} H.B.G. Casimir, Proc. Kon. Ned. Akad. Wet.
{\bf 51}, 793 (1948).
\smallskip
\item{6.} F. Wegner and A. Houghton, Phys. Rev. {\bf A 8},
401 (1973);\hfil\break
A. Hasenfratz and P. Hasenfratz, Nucl. Phys. {\bf B 270}, 685 (1986);
\hfil\break
C. Wetterich, Phys. Lett. {\bf B 301}, 90 (1993);\hfil\break
S.B. Liao and J. Polonyi, ``Renormalization group and universality'',
Duke TH-94-64.
\smallskip
\item{7.} C. Ford, D.R.T. Jones, P.W. Stephenson and
M.B. Einhorn, Nucl. Phys. {\bf B 395}, 17 (1993).
\smallskip
\item{8.} B. Warr, Ann. of Phys (NY) {\bf 183}, 1 (1988) \hfil\break
C. Becchi, ``On the construction of renormalized quantum field theory
using renormalization group techniques'', in {\it Elementary
Particles, Field Theory and Statistical Mechanics},
M. Bonini, G. Marchesini and E. Onofri, eds.,
University of Parma (1993);\hfil\break
M. Bonini, M. D'Attanasio and G. Marchesini, Nucl. Phys. {\bf B 418},
81 (1994); {\it ibid.} {\bf B 421},
429 (1994); ``BRS symmetry for Yang--Mills theory with exact
renormalization group'', Parma Preprint, UPRF 94-412 (1994).
\smallskip
\item{9.} M. Reuter and C. Wetterich, Nucl. Phys. {\bf B 408},
91 (1993); Nucl. Phys. {\bf B 417}, 181 (1994).
\smallskip

\vfil
\eject
\bye